# Hidden magnetic order on a kagome lattice for $KV_3Sb_5$

V. Scagnoli[1,2], D. D. Khalyavin[3] and S. W. Lovesey[3,4]

[1]Laboratory for Mesoscopic Systems, Department of Materials, ETH Zurich, 8093 Zurich, CH

[2]Laboratory for Multiscale Materials Experiments, Paul Scherrer Institute, 5232 Villigen PSI, CH

[3]ISIS Facility, STFC, Didcot, Oxfordshire OX11 0QX, UK

[4]Diamond Light Source Ltd, Didcot, Oxfordshire OX11 0DE, UK

**Abstract** $KV_3Sb_5$ has recently attracted a considerable attention, due to its low temperature superconducting properties, which are heralded by a charge density wave. The apparent presence of a very weak magnetism does not result in long range ordering. We propose a model compatible with a detectable internal magnetic field with no evidence of magnetic long-range order. It invokes higher order terms in the vanadium magnetization density compatible with the presence of a "hidden order" of Dirac (polar) multipoles. The Dirac dipole, known as an anapole or toroidal dipole, is one of a family of electronic multipoles visible in x-ray and magnetic neutron diffraction while undetectable with standard laboratory-based techniques. Two plausible hidden orders are studied with a view to testing in future x-ray and neutron Bragg diffraction experiments whether they are trustworthy. One candidate is magneto-electric and restricted to the linear type, while the other candidate cannot show a magneto-electric effect of any type. The latter hosts a vanadium entity that is both a true scalar and magnetic, and its presence epitomizes an absence of loop currents. Diffraction patterns for the two proposed hidden magnetic orders are distinctly different, fortunately. Corresponding scattering amplitudes for resonant x-ray and neutron Bragg diffraction are calculated with standard chemical and magnetic symmetry tools, and atomic physics.

## I. INTRODUCTION

High temperature superconductors are arguably amongst the most studied systems in solid state physics. Decades after its first observation in a Ba−La−Cu−O system, there is still no consensus on its microscopic origin [1, 2, 3]. One hoary question is the relevance of orbital currents, once predicted to be the origin of the hidden phase of high-temperature cuprate superconductors as well as the origin of the quantum anomalous Hall effect [4, 5, 6]. Such currents run through the lattice breaking time-reversal symmetry, and there are recent reports suggesting their presence in the kagome (trihexagonal tiling) superconductors $AV_3Sb_5$ (A = K, Rb, Cs) [7]. This family of compounds, which crystallizes in the hexagonal P6/mmm space group, has recently attracted a significant attention due the anomalies observes in its magnetic and electrical properties. At around a temperature T* ~ 100 K changes in the electrical resistivity and magnetic susceptibility are observed [8, 9]. Such anomalies have been associated with the development of a charge density wave, whose response under an applied magnetic field points to the presence of chirality in the system [7]. Additionally, $AV_3Sb_5$ becomes superconducting in the 0.9 - 2.5K temperature region [8, 10, 11]. Understandably, there are speculations on the exact nature of the electronic and magnetic state of $AV_3Sb_5$ below the temperature T* associated with the development of the charge density wave.

No magnetic order and concomitant time-reversal symmetry breaking has been revealed in $KV_3Sb_5$ to date. Powder neutron diffraction measurements find no obvious evidence of long-range or short-range magnetic ordering below 80 K [10]. In addition, Kenney *et al.*, find no evidence from muon spin relaxation and rotation spectroscopy (μSR) of a local vanadium ($V^{4+}$) magnetic moment [12]. However, weak internal magnetic fields are observed with zero-field μSR, and the authors speculate that it is due to simultaneous time-reversal and rotational symmetry breaking [13]. (Implanted muons occupy a variety of sites in a sample and the origin of detected magnetic fields is speculation.) Here, we show that these experimentally determined properties can be reconciled in a magnetic structure not visible with standard experimental techniques. Notably, Bragg diffraction patterns gathered with a polarized neutron beam, or x-rays tuned in energy to a vanadium atomic resonance can validate our magnetic models for $KV_3Sb_5$. (Neutron polarization analysis offers a significantly higher sensitivity to magnetic scattering than the standard powder diffraction technique.) Scattering amplitudes for neutron and x-ray diffraction presented here have been calculated with crystal and magnetic symmetry tools.

In more detail, we assign Dirac multipoles on vanadium ions to address the enigmatic magnetic properties of $KV_3Sb_5$. They are polar (parity-odd) and magnetic (time-odd) electronic multipoles, and arguably more profound physical objects than orbital currents, also known as current loops, and magnetic fields they are alleged to create [13]. That is to say, we propose abandoning space inversion symmetry while retaining violation of time reversal symmetry. Our proposed magnetic motifs are undistorted descendants of the kagome chemical structure. Hosting a vanadium entity that is both a true scalar and magnetic epitomizes an absence of current loops. Likewise in the case of ceramic superconductor materials, where copper Dirac multipoles furnish a good account of magnetism in the pseudo-gap phase. The value of the successful analysis of a neutron Bragg diffraction pattern from Hg1201 has been bolstered by a microscopic account of the formation of the relevant magnetic order parameter [14, 15, 16]. Specifically, anti-inversion symmetry in Cu sites emerges from centrosymmetric sites in the parent chemical structure. Moreover, the Dirac dipole depicted in Fig. 1, also called an anapole or toroidal dipole, has been directly observed in both neutron and x-ray diffraction [17, 18]. The work of Fernández-Rodríguez *et al.* [18] is particularly relevant to our current story for they deployed resonant x-ray diffraction to observe the vanadium anapole in vanadium sesquioxide ($V_2O_3$).

## II. SYMMETRY CONSIDERATIONS AND MAGNETIC MOTIFS

First, the chemical structure of the kagome metal. It incorporates a net of two regular tilings, one hexagonal and one triangular, that accommodate $V_3Sb_1$ with vanadium cations coordinated by octahedra of antinomy. Specifically, $KV_3Sb_5$ is described by the centrosymmetric space group P6/mmm (No. 191) with cell dimensions $a \approx 5.483$ Å, $c \approx 8.954$ Å [10, 13]. Ions occupy centrosymmetric sites, namely, $Sb_1$ 1(a) $D_{6h}$, $Sb_2$ 4(h) $C_{3v}$, V 3(f) $D_{2h}$ and K 1(b) $D_{6h}$. Vanadium ions originate the kagome network, while the $Sb_1$ ions fill the centres of the triangles. Also, $V_3Sb_1$ layers alternates with $Sb_2$ layers, resulting in a quasi-two-dimensional structure. Vectors describing the $KV_3Sb_5$ unit cell are **a** = ($a$, 0, 0), **b** = (1/2) (−$a$, $a\sqrt{3}$, 0) and **c** = (0, 0, $c$) in an orthonormal coordinate system.

Conventionally, the development of magnetic order leads to a lowering of the symmetry in the sample and the magnetic ordering pattern can be inferred by confronting experimental patterns derived by neutron diffraction (and to a minor extend also by x-ray magnetic diffraction) with a symmetry analysis in which selected elements of crystal symmetry are assumed to have disappeared. In the case of $KV_3Sb_5$, complete absence of conventional axial magnetic dipoles implies that any other magnetism is not visible using many experimental techniques - symmetry protects a magnetic order hidden from view. Following the experimental reports we have mentioned, we postulate that a presence of anti-inversion (the union of spatial and time reversals) forbids all axial magnetic moments in $KV_3Sb_5$ (anti-inversion occurs in 21 out of the 122 magnetic point groups). Our models of hidden magnetic order are derived from the chemical structure, with exclusive deployment of vanadium Dirac multipoles. Actually, symmetry in the two models of $KV_3Sb_5$ protects against axial magnetic dipoles at all occupied sites, and, consequently, the models respond like a non-magnetic material in usual experimental investigations in a laboratory environment. Furthermore, by stipulating that structures harbouring hidden order are direct derivatives of the chemical structure, we find there are just two candidates. By not seeking alternative hidden-order structures we have appealed to Occam's Razor.

Plausible magnetic motifs derived from the parent structure of $KV_3Sb_5$ include P6/m'mm and P6'/mmm' (Nos. 191.235 and 191.237 BNS [19]) selected for study. Ions in these two magnetic space groups do not possess conventional magnetism, with $Sb_1$, V and K ions in sites that include anti-inversion ($\bar{1}'$) and $Sb_2$ occupies sites with polar symmetry that prohibits axial dipoles. Spatial inversion is absent at all sites in question. Magnetic crystal classes 6/m'mm and 6'/mmm' are centrosymmetric, and not compatible with ferromagnetism. Any kind of magnetoelectric (ME) effect is prohibited by symmetry in 6'/mmm'. The magnetic motif hosts a new and novel entity that is a true scalar and time odd. We can associate it with a charge density and a fictitious electric field that satisfies Gauss's Law. A linear ME effect (EH type) is allowed by 6/m'mm but non-linear types are not symmetry allowed, e.g., EHH and EEH are forbidden where E (polar and time -even) and H (axial and time-odd) are electric and magnetic fields, respectively. Crystal chirality does not exist in the two models of $KV_3Sb_5$.

For our atomic description of electronic properties, vanadium ions are assigned spherical multipoles $\langle O^K_Q \rangle$ with integer rank K and projections Q in the interval $-K \leq Q \leq K$. Cartesian and spherical components of a dipole **R** = (x, y, z), for example, are related by x = $(R_{-1} - R_{+1})/\sqrt{2}$, y = $i(R_{-1} + R_{+1})/\sqrt{2}$, z = $R_0$ [21]. A complex conjugate is defined as $\langle O^K_Q \rangle^* = (-1)^Q \langle O^K_{-Q} \rangle$, and a phase convention $\langle O^K_Q \rangle = [\langle O^K_Q \rangle' + i \langle O^K_Q \rangle'']$ for real and imaginary parts labelled by single and double primes, respectively. In which case, the diagonal multipole $\langle O^K_0 \rangle$ is purely real. Angular brackets $\langle ... \rangle$ denote the time-average, or expectation value, of the enclosed tensor operator, i.e., vanadium multipoles feature in the electronic ground state of $KV_3Sb_5$.

To establish the experimental conditions required to validate our predictions we present amplitudes, or unit-cell structure factors, for resonant x-ray and magnetic neutron diffraction derived from an electronic structure factor [21]. Its generic form is identical for vanadium ions at Wyckoff position 3(f) in P6/m'mm and P6'/mmm', as we see in Eq. (A2). Different site symmetries alone account for differences in diffraction amplitudes.

Our starting point is the selection of the most appropriate reference frame to describe the symmetry properties of the V ions. Reciprocal lattice vectors are found to be $\mathbf{a}^* = (ac/2)(\sqrt{3}, 1, 0) \propto [2, 1, 0]$, $\mathbf{b}^* = (ac)(0, 1, 0) \propto [1, 2, 0]$ and $\mathbf{c}^* = \sqrt{3}(a^2/2)(0, 0, 1)$ all in units of $(2\pi)/v_o$ with a volume $v_o = \sqrt{3}(a^2c/2)$. Miller indices $(h, k, l)$ are integers. Our local axes for vanadium ions labelled $(\xi, \eta, \zeta)$ match orthogonal vectors $\mathbf{a}$, $\mathbf{b}^*$ and $\mathbf{c}$. An anti-dyad $2'_\xi$ and anti-inversion are 3(f) site symmetries in both candidate structures. Remaining site symmetries are $2'_\eta$, $2_\zeta$ and $2_\eta$, $2'_\zeta$ for P6/m'mm and P6'/mmm', respectively. X-ray and neutron diffraction amplitudes for reflection vectors $(h, 0, 0)$ and $(0, 0, l)$ are very different in the two candidates, as we will see. Bragg angles are determined by $\sin(\theta) \approx (h\, 0.239)$ and $\sin(\theta) \approx (l\, 0.127)$.

### III. RESONANT X-RAY DIFFRACTION

Valence states that accept the photo-ejected electron, a few eV above the Fermi level, interact with neighbouring ions. In consequence, any corresponding electronic multipole is rotationally anisotropic with a symmetry corresponding to the site symmetry of the resonant ion. This anisotropy is most pronounced in the direct vicinity of an absorption edge whereas it is negligible far from the edges. Non-resonant ions can be neglected in calculations of forbidden reflection structure factors, to a good approximation. There are many reported examples of Bragg diffraction enhanced by absorption at the K-edge of a 3d transition ion. Results on haematite ($Fe^{3+}$, $3d^5$) reported by Finkelstein *et al*. [22] are thoroughly discussed by Carra and Thole [23], while diffraction patterns gathered at a later date revealed a chirality [24]. The time between the publications saw reports of diffraction patterns enhanced by nickel and vanadium K-edges [18, 25-27].

Absorption at the K edge and an electric-dipole event (E1) gives access to valence states with atomic p-like character, and an electric quadrupole event (E2) at the same edge gives access to d-like states (1s → 3d). Bragg diffraction from $V_2O_3$ x-ray enhanced by the pre-edge ≈ 5.464 keV (wavelength $\lambda \approx 2.27$ Å) feature of the vanadium K shell has been successfully analysed using the electric dipole – electric quadrupole (E1-E2) absorption event [18, 28]. The triangle rule says that multipoles possess ranks K = 1, 2, 3 for this parity-odd event [21, 29]. Dipoles (K = 1) in the parity-even E1-E1 absorption event are magnetic, and irrelevant to our study of $KV_3Sb_5$. First, parity-even (axial) magnetic multipoles do not exist at sites that contain anti-inversion, and such is the case for $Sb_1$, V and K ions. Secondly, axial dipoles do not exist at sites 4(h) used by $Sb_2$. However, a parity-even quadrupole and octupoles (K = 3) with projections Q = ±3 are permitted by 4(h) symmetry. The Sb multipoles would contribute to diffraction enhanced by a Sb E1-E1 or E2-E2 absorption events but it does not exist at the vanadium K-edge of interest here. In summary, vanadium diffraction patterns presented below exist with broken time-reversal symmetry, polar magnetism and concomitant vanadium Dirac multipoles. We focus on x-ray amplitudes in which polarization is rotated, e.g., $\sigma \to \pi'$ in Fig. 2, for Thomson scattering is absent in the rotated channel of polarization. In the nominal setting of the crystal $(\xi, \eta, \zeta)$ coincide with $(x, y, z)$ in Fig. 2, which depicts four states of x-ray polarization.

A resonant atomic process may dominate all other contributions to the x-ray scattering length should the photon energy E match a resonance energy $\Delta$. Assuming virtual intermediate states in the process are spherically symmetric, to a good approximation, the x-ray scattering length ≈ {(μη)/(E − Δ + iΓ/2)} in the region of the resonance, where Γ is the total width of the resonance [21]. The numerator (μη) is an amplitude, or unit-cell structure factor, for Bragg diffraction in the scattering channel with primary (secondary) polarization η (μ). By convention, σ denotes polarization normal to the plane of scattering, and π denotes polarization within the plane of scattering, as in Fig. 2. The illuminated crystal is rotated about the reflection vector in an azimuthal angle scan. Intensity of a Bragg spot in the σ → π' channel of polarization is proportional to |(π′σ)|$^2$, and likewise for unrotated channels of polarization.

A signature of crystal chirality is the difference between Bragg intensities measured with left- and right-handed primary x-rays, or x-rays with opposite helicities. A relevant quantity, ϒ, depends on all four scattering amplitudes, ϒ = {(σ′π)*(σ′σ) + (π′π)*(π′σ)}″. Specifically, crystal chirality means an intensity $P_2$ϒ different from zero, where the Stokes parameter $P_2$ (a purely real pseudo-scalar) measures helicity in the primary x-ray beam. Since intensity is a true scalar, ϒ and $P_2$ must possess identical discrete symmetries, specifically, both scalars are time-even and parity-odd. The chiral signature ϒ for vanadium ions is zero for all Bragg diffraction patterns that we choose to discuss.

In the following, we give explicit expression for the contributions to the scattering cross section for the two candidate structures: A) P6'/mmm' with any kind of magnetoelectric (ME) effect is prohibited by symmetry in 6'/mmm', and B) P6/m'mm with a linear ME effect permitted by 6/m'mm. All reported x-ray scattering amplitudes are derived from universal expressions for diffraction by Dirac multipoles visible in x-ray scattering enhanced by an E1-E2 absorption event [29].

### A. P6'/mmm' (non magnetoelectric)

In this model, sites 3(f) used by V ions possess symmetry mmm'. Projections Q on Dirac multipoles ⟨$G^K_Q$⟩ are odd by virtue of the anti-dyad 2'$_ζ$ in the symmetry elements of the point group, and, consequently, magnetic charge ⟨$G^0$⟩ is not allowed. In addition, symmetry 2'$_ξ$ is satisfied by ⟨$G^K_{-Q}$⟩ = (−1)$^{K + 1}$ ⟨$G^K_Q$⟩ that leads to the identity ⟨$G^K_Q$⟩* = (−1)$^K$ ⟨$G^K_Q$⟩ for Q odd. Anapoles ⟨$G^1$⟩ depicted in Fig. 3 are parallel to the reciprocal vector **b**\*, i.e., the local η axis.

Inspection of Fig. 3 reveals an unusual entity at the midpoint of each triangle, with antimony ions displaced above and below the plane of the triangle. The entity is a true scalar and time odd. (Whereas, a like entity created with axial dipoles is a Dirac monopole that can be accommodated in Maxwell's equations of electrodynamics.) We equate the entity with $i$ρ, where ρ is a purely real classical density. This line of argument uses the fact that time reversal includes the operation of complex conjugation [30]. Upon associating ρ with electric charge a conjugate (fictitious) field **E** satisfies div **E** = ρ (Gauss's Law), and its magnitude as a function of distance obeys an inverse square law.

Evaluated for (0, 0, $l$), the unit-cell structure factor Eq. (A2) is proportional to [1 + 2 cos($2\pi Q/3$)] that is zero for projections $Q = \pm 1$. Projections $Q = \pm 3$ are allowed, however, and can only belong to an octupole ($K = 3$). Amplitudes of diffraction in rotated channels of polarization ($\pi'\sigma$) and ($\sigma'\pi$), and a Bragg spot (0, 0, $l$) are,

$$(\pi'\sigma) = - (\sigma'\pi) = 3 \sin(2\theta) \sin(3\psi) \langle G^3_{+3}\rangle''. \quad (0, 0, l) \qquad (1)$$

Here, $\theta$ is the Bragg angle depicted in Fig. 2, and $\psi$ is the angle of rotation of the crystal around the reflection vector (0, 0, $l$). At the azimuthal origin $\psi = 0$ the crystal **a** axis is normal to the plane of scattering in Fig. 2. Notable features of amplitudes in Eq. (1) include a three-fold periodicity in $\psi$ from the tertiary axis of rotation symmetry, sole dependence on a lone octupole, and handedness observed in the sign difference between ($\pi'\sigma$) and ($\sigma'\pi$). Amplitudes in unrotated channels of polarization are significantly different with respect to $\psi$. For example,

$$(\sigma'\sigma) = - (3/2) \cos(\theta) [19 \cos(3\psi) + \cos(\psi)] \langle G^3_{+3}\rangle'' . \quad (0, 0, l) \qquad (2)$$

Amplitudes ($\sigma'\sigma$) and ($\pi'\sigma$) have opposite trends with respect to increasing $l$, and are even and odd functions of $\psi$, respectively.

Amplitudes ($\sigma'\sigma$) and ($\pi'\pi$) for a reflection vector ($h$, 0, 0) are zero. Amplitudes ($\pi'\sigma$) and ($\sigma'\pi$) can be non-zero, however, and they are different for $h$ odd and $h$ even. First, $h$ even amplitudes depend on the octupole visible in (0, 0, $l$) Bragg spots, namely,

$$(\pi'\sigma) = (\sigma'\pi) = - 3 \cos^2(\theta) \sin(2\psi) \langle G^3_{+3}\rangle''. \quad (2n, 0, 0) \qquad (3)$$

The crystal **c** axis is normal to the plane of scattering for $\psi = 0$. Amplitudes for $h$ odd engage another three multipoles yet possess the dependence on $\theta$ and $\psi$ displayed in Eq. (3). Additional multipoles are $\langle G^2_{+1}\rangle'$, $\langle G^3_{+1}\rangle''$ and $\langle G^3_{+3}\rangle''$.

### B. P6/m'mm (magnetoelectric)

Sites 3(f) occupied by V ions possess symmetry m'mm. Projections on Dirac multipoles $\langle G^K_Q\rangle$ are even by virtue of the dyad $2\zeta$, meaning $Q = 0, \pm 2$ for an E1-E2 absorption event with $K = 1, 2, 3$. The anapole $\langle G^1_0\rangle$ is parallel to the hexagonal axis **c**, and the ferro-motif is depicted in Fig. 4. Results $\langle G^K_{-Q}\rangle = (-1)^{K+1} \langle G^K_Q\rangle$ and $\langle G^K_Q\rangle^* = - (-1)^K \langle G^K_Q\rangle$ follow from the symmetry element 2'$\xi$. Like the non-ME motif, magnetic charge $\langle G^0\rangle$ is prohibited. All P6/m'mm scattering amplitudes are zero for Bragg spots (0, 0, $l$), however, while like spots offer a direct test on the existence of $\langle G^3_{+3}\rangle''$ in the other candidate structure.

Turning to ($h$, 0, 0) and $h$ even,

$$(\pi'\sigma) = - (\sigma'\pi) = \sin(2\theta) \cos(\psi) [\sqrt{6} \langle G^1_0\rangle - (5 \cos(2\psi) - 3) \langle G^3_0\rangle]. \quad (2n, 0, 0) \qquad (4)$$

Notably, only diagonal multipoles (Q = 0) contribute, including the anapole parallel to the crystal **c** axis. With regard to a dependence on the azimuthal angle, cos($\psi$) identifies [$\langle G^1_0 \rangle$ + $\sqrt{(3/2)}$ $\langle G^3_0 \rangle$] and cos($\psi$)cos($2\psi$) identifies $\langle G^3_0 \rangle$, i.e., the two contributing multipoles can be separated in an analysis of experimental data. Similar results are obtained for unrotated amplitudes, e.g.,

$$(\sigma'\sigma) = -2 \cos(\theta) \sin(\psi) [\sqrt{6} \langle G^1_0 \rangle + (5 \cos(2\psi) + 3) \langle G^3_0 \rangle].(2n, 0, 0) \qquad (5)$$

There are an additional two multipoles in amplitudes with *h* odd. Specifically,

$$(\pi'\sigma) = -(\sigma'\pi) = (2/3) \sqrt{(5/3)} \sin(2\theta) \cos(\psi) \{-\sqrt{(3/20)} [\sqrt{6} \langle G^1_0 \rangle - (5 \cos(2\psi) - 3) \langle G^3_0 \rangle]$$

$$- 4 \langle G^2_{+2} \rangle'' + \sqrt{2} (3 \cos(2\psi) - 1) \langle G^3_{+2} \rangle'\}. (2n + 1, 0, 0) \qquad (6)$$

The ($\sigma'\sigma$) amplitude is composed of the same multipoles, and cos($\theta$) sin($\psi$) multiplies and even function of $\psi$. Thus, for *h* = 2*n* + 1, amplitudes ($\sigma'\sigma$) and ($\pi'\sigma$) are odd and even functions of $\psi$, respectively.

## IV. MAGNETIC NEUTRON DIFFRACTION

A magnetic scattering amplitude $\langle \mathbf{Q}_\perp \rangle$ generates an intensity $|\langle \mathbf{Q}_\perp \rangle|^2$ of unpolarized neutrons. Dirac multipoles contribute to scattering, and each one is accompanied by an atomic form factor that depends on the magnitude of the reflection vector and the electronic configuration [31].

Polarization analysis measures the magnetic content of a Bragg spot with overlapping nuclear and magnetic amplitudes, which occurs when the magnetic motif and chemical structure coincide [16, 20]. Primary and secondary polarizations are denoted **P** and **P′**, and a fraction (1 − **P** • **P′**)/2 of neutrons participate in events that change (flip) the neutron spin orientation. For a collinear magnetic motif one finds (1 − **P** • **P′**)/2 $\propto$ {(1/2) (1 + $P^2$) $|\langle \mathbf{Q}_\perp \rangle|^2$ − $|\mathbf{P} \cdot \langle \mathbf{Q}_\perp \rangle|^2$}. A quantity called spin-flip,

$$(SF) = \{|\langle \mathbf{Q}_\perp \rangle|^2 - |\mathbf{P} \cdot \langle \mathbf{Q}_\perp \rangle|^2\}, \qquad (7)$$

obtained with $P^2 = 1$ is a standard measure of the strength of magnetic scattering [16].

A Dirac dipole $\langle \mathbf{D} \rangle$ in neutron diffraction is the sum of three contributions that include expectation values of spin and orbital anapoles. Electronic spin, orbital angular momentum and position operators are **S**, **L** and **R**, respectively. Operators for the three contributions to **D** are a spin anapole (**S**×**R**), orbital anapole $\mathbf{\Omega} = [\mathbf{L}\times\mathbf{R} - \mathbf{R}\times\mathbf{L}]$, and (*i***R**). Specifically, $\langle \mathbf{D} \rangle$ = (1/2) [3 ($h_1$) $\langle (\mathbf{S}\times\mathbf{R}) \rangle$ − ($j_0$) $\langle \mathbf{\Omega} \rangle$ + ($g_1$) $\langle (i\mathbf{R}) \rangle$]. Form factors ($h_1$), ($j_0$) and ($g_1$) have been calculated for several atomic configurations [32, 33]. In what follows, we retain $\langle \mathbf{D} \rangle$ and a quadrupole

⟨$H^2$⟩ that possess the largest atomic form factors. A quadrupole of this type is a product of (h$_1$) and a correlation function ⟨{**S** ⊗ **R**}$^2$⟩ written in terms of a standard tensor product of two dipoles [31]. Notably, ⟨$H^2$⟩ ∝ [(h$_1$) ⟨{**S** ⊗ **R**}$^2$⟩] accounts for magnetic neutron diffraction by the pseudo-gap phase of ceramic superconductors [20].

### A. P6'/mmm' (non magnetoelectric)

The motif of anapoles is depicted in Fig. 3. No diffraction takes place with anapoles and quadrupoles at Bragg spots indexed by (0, 0, $l$). Likewise, for ($h$, 0, 0) with $h$ = 2$n$. Amplitudes for ($h$, 0, 0) with Miller index $h$ odd are, ⟨$Q_{\perp\xi}$⟩ ≈ 0, ⟨$Q_{\perp\eta}$⟩ ≈ 0 and,

⟨$Q_{\perp\zeta}$⟩ ≈ $i$[− ⟨$D_\eta$⟩ + (3/√5) ⟨$H^2_{+1}$⟩']. (2$n$ + 1, 0, 0)  (8)

The dipole contribution is parallel to the reciprocal vector **b**\*. From Eqs. (7) and (8), (SF) = 0 for polarization **P** parallel to the crystal **c** axis, while (SF) = |⟨**Q**$_\perp$⟩|$^2$ for **P** in the basal plane.

### B. P6/m'mm (magnetoelectric)

As with the motif depicted in Fig. 3, there is no diffraction at Bragg spots indexed by (0, 0, $l$) in the ferro-type motif depicted in Fig. 4. However, diffraction at Bragg spots ($h$, 0, 0) occurs at both $h$ even and $h$ odd. One finds, ⟨$Q_{\perp\xi}$⟩ ≈ 0, ⟨$Q_{\perp\zeta}$⟩ ≈ 0 and,

⟨$Q_{\perp\eta}$⟩ ≈ $i$ [⟨$D_\zeta$⟩ − (3/√5) ⟨$H^2_{+2}$⟩"], (2$n$ + 1, 0, 0)  (9)

for $h$ odd. The quadrupole in Eq. (9) is absent for $h$ = 2$n$. Evidently, (SF) = |⟨**Q**$_\perp$⟩|$^2$ for **P** parallel to either [2, 1, 0] or [0, 0, 1] crystal axes, while (SF) = 0 for **P** parallel to the crystal **b**\* axis.

### V. CONCLUSIONS

In summary, we have studied two models of a magnetic material on a kagome lattice with a view to interpreting enigmatic properties of KV$_3$Sb$_5$ at a low temperature. Their distinguishing features are a complete absence of conventional (axial, parity-even) magnetic dipoles, and a lattice structure that is a direct descendent of the kagome chemical structure. The latter feature is evident in the specification of the models by magnetic space groups P6'/mmm' and P6/m'mm derived from the kagome structure P6/mmm. Anapole motifs are shown in Figs. 3 and 4. Magnetic properties of the models chime with the absence of magnetic long-range order and, also, a vanadium magnetic moment in available measurements [10, 12].

Symmetry in the form of anti-inversion, the union of space and time inversions, prohibits all axial vanadium magnetic multipoles. Magnetism exists in the form of polar magnetic multipoles. The prediction of a magnetic true scalar (time-odd and parity-even) in the kagome lattice is a vivid portrayal of the unconventional nature of the magnetism. Notably, the discrete symmetries of the scalar are not those of vanadium Dirac multipoles (time-odd and parity-odd). Dirac multipoles deflect neutrons and x-rays so future Bragg diffraction experiments using these radiations can inform us whether the models we have studied are noteworthy. To this end, scattering amplitudes for neutron diffraction, and x-ray diffraction with signal enhancement by a vanadium atomic resonance are reported. Resonant x-ray Bragg diffraction by vanadium Dirac dipoles, also called anapoles (Fig. 1), has previously been identified in the diffraction pattern of $V_2O_3$ [18]. In addition, there is direct evidence of neutron scattering by anapoles [17].

Our two models possess different magneto-electric (ME) properties. Any type of ME effect is prohibited in P6'/mmm', whereas a linear ME effect is permitted in P6/m'mm. Vanadium magnetic monopoles (Dirac scalars) are forbidden in both models. Anapoles are aligned with **b*** (**c**) in P6'/mmm' (P6/m'mm), where the reciprocal lattice vector **b*** is orthogonal to lattice vectors **a** and **c** in Figs. 3 and 4. As an example of how neutron Bragg diffraction can be used to differentiate between the two models we mention the spin-flip signal reported in Section IV. In brief, the signal is zero for neutron polarization parallel to the crystal **b*** (**c**) axis for P6/m'mm (P6'/mmm'). Regarding resonant x-ray diffraction with signal enhancement by an electric dipole - electric quadrupole (E1-E2) event at the vanadium K-edge, Bragg spots are allowed for a reflection vector parallel to the c-axis in P6'/mmm', while no such Bragg spots are allowed in P6/m'mm. Section III includes scattering amplitudes for a reflection vector in the basal plane.

Sites 4(h) used by $Sb_2$ in our models do not contain anti-inversion or inversion symmetries. In the absence of anti-inversion conventional magnetism (parity $\sigma_\pi = +1$) is not forbidden, unlike all other sites in our two models of $KV_3Sb_5$, and might arise from back-electron transfer. For completeness, we survey the multipoles $\langle O^K_Q \rangle$ of rank K allowed by site symmetries. For sites 4(h) they are $3\zeta$, $m_\xi = I2_\xi$ and $\langle O^K_Q \rangle = (-1)^Q \exp(-i\pi Q/3) [m_\xi \langle O^K_Q \rangle]$ leading to projections Q = 3n and $\langle O^K_Q \rangle = \sigma_\pi (-1)^{K+Q} \langle O^K_Q \rangle^*$. Orthogonal crystal vectors **a**, **b*** and **c** define ($\xi$, $\eta$, $\zeta$). Allowed parity-even magnetism includes multipoles $\langle T^2_0 \rangle$ and $\langle T^3_{+3} \rangle'$, and associated magnetic fields might be sensed in zero-field µSR. Both multipoles contribute to neutron scattering, while just $\langle T^3_{+3} \rangle'$ can be seen in a parity-even absorption event, namely, E2-E2. Dirac multipoles permitted in neutron scattering and an E1-E2 absorption event include an anapole $\langle G^1_0 \rangle$ and octupoles $\langle G^3_0 \rangle$ and $\langle G^3_{+3} \rangle''$.

Anti-inversion symmetry at sites used by vanadium ions is the key component of our filter of magnetic models that descend from the kagome lattice. Absence by symmetry of axial magnetic dipoles at all sites used by the three elements in $KV_3Sb_5$ is a further requirement. The second component eliminates P6'/mm'm (No. 191.236 [19]) and P6/m'm'm' (No. 191.241) as acceptable models, because magnetic dipoles are permitted at sites 4(h). According to conventional displays of magnetic space groups our two magnetic models of $KV_3Sb_5$ are actually classified as non-magnetic.

## APPENDIX

An electronic structure factor,

$$\Psi^K_Q = [\exp(i\boldsymbol{\kappa} \cdot \mathbf{d}) \langle O^K_Q \rangle_\mathbf{d}], \quad (A1)$$

where the reflection vector **κ** is defined by Miller indices (h, k, l), and the implied sum is over three vanadium ions in sites **d**, 3(f). For candidates P6/m'mm and P6'/mmm' (Nos. 191.235 and 191.237 [19]),

$$\Psi^K{}_Q(3f) = \langle O^K{}_Q\rangle\,[(-1)^h + (-1)^{h+k}\,\gamma_Q{}^* + (-1)^k\,\gamma_Q], \tag{A2}$$

with $\gamma_Q = \exp(i2\pi Q/3)$ and reference site (1/2, 0, 0) for $\langle O^K{}_Q\rangle$. Our two candidates differ with respect to site symmetries that delineate properties of $\langle O^K{}_Q\rangle$, namely, mmm' (No. 191.237) or m'mm (No. 191.235).

Scattering amplitudes in Refs. [29, 31] used here are functions of two quantities, one even and one odd with respect to the sign of Q. For neutron scattering $A^K{}_Q \pm B^K{}_Q = \Psi^K{}_{\pm Q}$ [31]. In the case of x-ray diffraction, it is necessary to align the reflection vector **κ** with the − x axis depicted in Fig. 2. For reflections (*h*, 0, 0), the result is $A^K{}_Q \pm B^K{}_Q = \exp(\pm iQ\alpha)\,\Psi^K{}_{\pm Q}$ with angle $\alpha = -30^\circ$. The corresponding result for a reflection vector (0, 0, *l*) is more complicated and it can be found in Eq. (104) in Ref. [21].

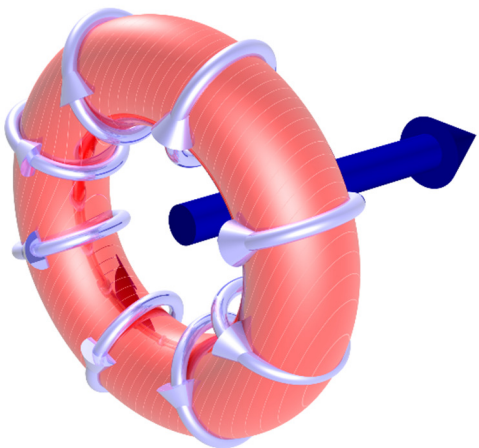

FIG. 1. Depiction of a toroidal dipole, also known as an anapole.

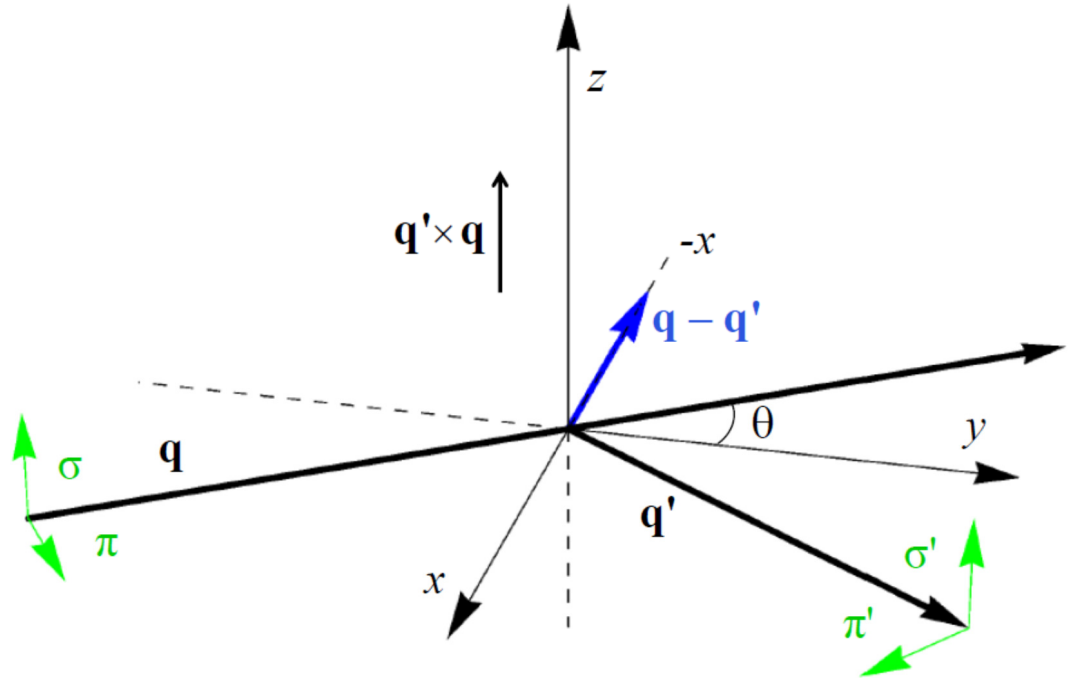

FIG. 2. Primary (σ, π) and secondary (σ′, π′) states of polarization. Corresponding wave vectors **q** and **q′** subtend an angle 2θ. The Bragg condition for diffraction is met when **q − q′** coincides with the reflection vector indexed ($h$, $k$, $l$). Crystal vectors **a**, **b*** and **c** that define (ξ, η, ζ) and depicted Cartesian (x, y, z) coincide in the nominal setting of the crystal.

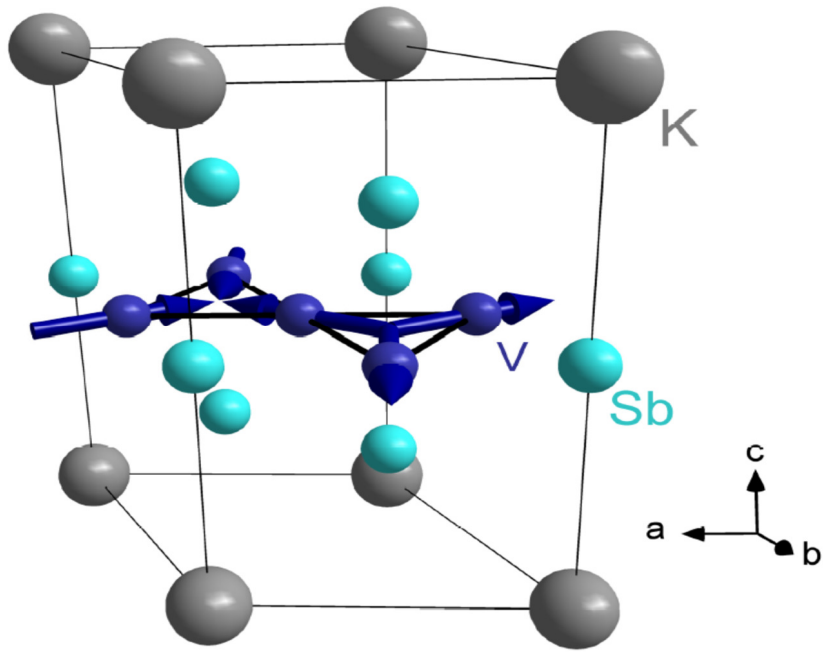

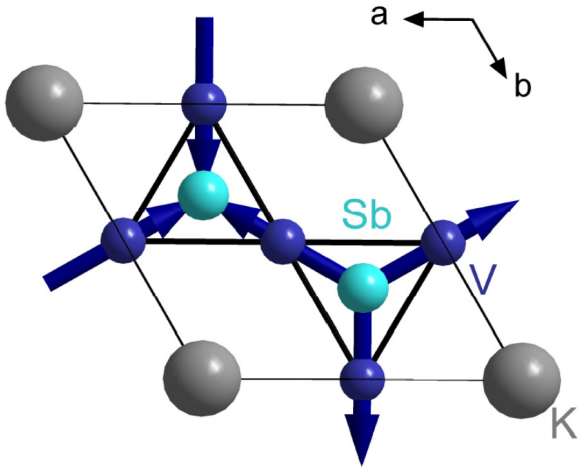

FIG. 3. Anapole motif in P6'/mmm' (No. 1911. 237) viewed in two perspectives. They are depicted in Fig. 1 and are represented here by dark blue arrows. Two anapoles are normal to the crystal **a** axis and parallel to hexagonal [1, 2, 0] ∝ **b**$^*$ in the bottom panel. Cell vectors **a** and **b** subtend an angle 120º (Section II). Light blue and grey spheres represent Sb and K ions, respectively.

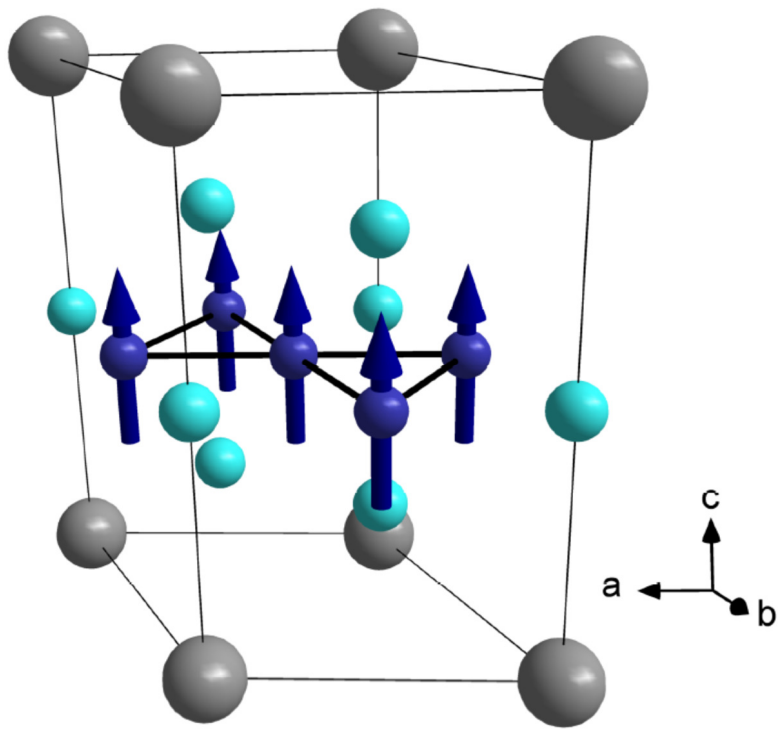

FIG. 4. Ferro-motif of anapoles in P6/m'mm (No. 191.235). Cell vectors **a** and **b** subtend an angle 120º (Section II). Light blue and grey spheres represent antimony and potassium ions, respectively.